\documentclass{IEEEtran}
\usepackage{graphicx,color}
\usepackage{subfig}
\usepackage{amsmath, epsfig, amssymb}
\usepackage{times,amsmath,epsfig,amssymb,graphicx,amsfonts}
\usepackage{algorithmic}
\usepackage[titlenotnumbered ,ruled]{algorithm2e}
\title{Optimality of Operator-Like Wavelets\\for Representing Sparse AR(1) Processes}

\author{Pedram Pad and Michael Unser\\Biomedical Imaging Group, EPFL, Switzerland\thanks{This work was supported by the European Commission under Grant ERC-2010-AdG 267439- FUN-SP.}}

\newcommand{\Indic}{\boldsymbol {1}}
\newcommand{\drm}{\mathrm{d}}

\newcommand{\ds}{\mathrm{d}s}

\newcommand{\e}{\mathrm{e}}

\newcommand{\Abold}{\mathbf{A}}
\newcommand{\Hbold}{\mathbf{H}}
\newcommand{\Lbold}{\mathbf{L}}
\newcommand{\Ubold}{\mathbf{U}}
\newcommand{\Lambdabold}{\mathbf{\Lambda}}
\newcommand{\Vbold}{\mathbf{V}}
\newcommand{\Kbold}{\mathbf{K}}
\newcommand{\Ibold}{\mathbf{I}}
\newcommand{\Dbb}{\mathbb{D}}
\newcommand{\Dop}{\mathrm{D}}
\newcommand{\Iop}{\mathrm{I}}
\newcommand{\hdot}{\bar{h}}
\newcommand{\Hbb}{\mathbb{H}}
\newcommand{\Lop}{\mathrm{L}}
\newcommand{\sbold}{\mathbf{s}}

\newcommand{\ybold}{\mathbf{y}}
\newcommand{\zbold}{\mathbf{z}}
\newcommand{\nbold}{\mathbf{n}}
\newcommand{\nablabold}{\mbox{\boldmath$\nabla$}}

\newtheorem{note}{Note}
\newtheorem{theorem}{Theorem}
\newtheorem{property}{Property}

\begin{document}
\maketitle
\begin{abstract}
It is known that the Karhunen-Lo{\`{e}}ve transform (KLT) of {\emph{Gaussian}} first-order auto-regressive (AR(1)) processes results in sinusoidal basis functions. The same sinusoidal bases come out of the independent-component analysis (ICA) and actually correspond to processes with completely independent samples. In this paper, we relax the Gaussian hypothesis and study how orthogonal transforms decouple symmetric-alpha-stable (S$\alpha$S) AR(1) processes. The Gaussian case is not sparse and corresponds to $\alpha=2$, while $0<\alpha<2$ yields processes with sparse linear-prediction error. In the presence of sparsity, we show that operator-like wavelet bases do outperform the sinusoidal ones. Also, we observe that, for processes with very sparse increments ($0<\alpha\leq 1$), the operator-like wavelet basis is indistinguishable from the ICA solution obtained through numerical optimization. We consider two criteria for independence. The first is the Kullback-Leibler divergence between the joint probability density function (pdf) of the original signal and the product of the marginals in the transformed domain. The second is a divergence between the joint pdf of the original signal and the product of the marginals in the transformed domain, which is based on Stein's formula for the mean-square estimation error in additive Gaussian noise. Our framework then offers a unified view that encompasses the discrete cosine transform (known to be asymptotically optimal for $\alpha=2$) and Haar-like wavelets (for which we achieve optimality for $0<\alpha\leq1$).
\end{abstract}

\section{Introduction}
Transform-domain processing is a classical approach to compress signals, model data, and extract features. The underlying idea is that the transform-domain coefficients exhibit a loosened interdependence so that a simple point-wise processing can be applied. For instance, in the discrete domain, the Karhunen-Lo{\`{e}}ve transform (KLT) is famous for yielding optimal transform coefficients that are uncorrelated and therefore also independent, provided the process is Gaussian. Also, if the process is stationary with finite variance and infinite length, then the KLT is a Fourier-like transform (FT-like). Moreover, it is known that FT-like transforms such as the discrete cosine transform are asymptotically equivalent to the KLT for AR(1) processes~\cite{Pearl1973,Unser1984}; thus, for a Gaussian input, all these transforms result in a fully decoupled (independent) representation. However, this favorable independence-related property is extinguished for non-Gaussian processes. In this case, the coefficients are only partially decoupled and the representation of the signal afforded by the KLT is suboptimal.

In recent years, wavelets have emerged as an alternative representation of signals and images. Typical examples of successful applications are JPEG2000 for image compression~\cite{Taubman-etal2001} and shrinkage methods for attenuating noise~\cite{Donoho1993,Taswell2000}. Their wide success for transform-domain processing recommends them as good candidates for decoupling practical processes. This empirical observation was established by early studies that include~\cite{Olshausen-etal1996}, where many natural images were subjected to an independent-component analysis (ICA). It was found that the resulting components have properties that are reminiscent of 2D wavelets and/or Gabor functions. Additional ICA experiments were performed in~\cite{Cardoso-etal1999} on realizations of the stationary sawtooth process and of Meyer's ramp process~\cite{Meyer1992}; for both processes, the basis vectors of ICA exhibit a wavelet-like multiresolution structure.

Unfortunately, despite their empirical usefulness, the optimality of wavelets for the representation of non-Gaussian stochastic processes remains poorly understood from a theoretical point of view. An early study can be traced back to~\cite{Flandrin1989}, where the decomposition of fractional Brownian motions over a wavelet basis was shown to result in almost uncorrelated coefficients, under some conditions. (Ultimately however, the truly optimal domain to represent fractional Brownian motions is known to be simply the FT domain.) Meanwhile, in a deterministic framework, it was shown in~\cite{Devore1998} that wavelets are optimal (up to some constant) for the $N$-term approximation of functions in Besov spaces; the extension of this result to a statistical framework could be achieved only experimentally.

Recently, a new tool for the study of generalized innovation models has been proposed in~\cite{Unser-etal2011a,Unser-etal2011b}. It is particularly well suited to the investigation of symmetric-$\alpha$-stable (S$\alpha$S) white noises, which can be used to drive first-order stochastic differential equations (SDE) to synthesize AR(1) processes. As it turns out, AR(1) systems and $\alpha$-stable distributions are at the core of signal modeling and probability theory. The classical Gaussian processes correspond to $\alpha=2$, while $0<\alpha<2$ yields stable processes that have heavy-tailed statistics and that are prototypical representatives for sparse signals~\cite{Amini-etal2011}.

In this paper, we take advantage of this tool to establish the optimality of a certain class of wavelets in a stochastic sense. We start by characterizing the amount of dependency between the coefficients of stochastic processes represented in an arbitrary transform domain. We consider two measures of dependency. The first is based on the Kullback-Leibler divergence and the second is based on Stein's formula for the variance of estimation of a signal distorted by additive white Gaussian noise (AWGN). Then, we seek the orthogonal transformation that minimizes these measures of dependency. We confirm the extinction of the optimality of FT-like transforms for $0<\alpha<2$ and validate the superiority of the operator-like wavelet transforms proposed in~\cite{Khalidov-etal2006}. Also, by finding the optimal transform for different values of $\alpha$, we demonstrate that, for a positive $\alpha$ less than some threshold, operator-like wavelets are optimal.

This paper is organized as follows: We start by exposing three preliminary concepts like i) measures of divergence between distributions that would be suitable for either noise attenuation or compression applications (Section \ref{sec:permea}); ii) the signal model fundamental to this paper (Section \ref{subsec:sasproc}); and iii) operator-like wavelets, (Section \ref{subsec:OpWT}). In Section~\ref{sec:main}, we describe our performance criteria in the context of transform-domain compression and noise attenuation. In addition, we provide an iterative algorithm to find the optimal basis. Results for different AR(1) processes and different transform domains are discussed in Section~\ref{sec:results}. The last section is dedicated to the recapitulation of the main results, the relation to prior works, and topics for future studies.

\section{Performance Measures}\label{sec:permea}
In statistical modeling, one interesting problem is the best-achievable performance when the model does not match the reality. In the following we address this issue for the two problems of compression and denoising when the assumed distribution and the real one may differ.

\subsubsection{Compression Based on Non-Exact Distribution}\label{sec:ApCom}
It is well-known that, if we have a source $\sbold$ of random vectors with pdf $p_{\sbold}$, then the minimum coding set log-measure (MCSM) of these vectors is
\begin{eqnarray}\label{eq:mnb}
\text{MCSM}=\Hbb(p_\sbold)=-\int{p_\sbold(\sbold)\log p_\sbold(\sbold)\drm\sbold}
\end{eqnarray}
which is the entropy of the source. However, if we compress $\sbold$ assuming $q_{\sbold}$ as its distribution, then 
\begin{eqnarray}\label{eq:nb}
\text{CSM}(q_{\sbold})&=&\text{MCSM}+\Dbb(p_{\sbold}\|q_{\sbold})\nonumber\\&=&\text{MCSM}+\int{p_{\sbold}(\sbold)\log\frac{p_{\sbold}(\sbold)}{q_{\sbold}(\sbold)}\drm\sbold}
\end{eqnarray}
in which $\Dbb(\cdot\|\cdot)$ is the Kullback-Leibler divergence.

Typically, when there is a statistical dependency between the entries of $\sbold$, compressing the vector based on the exact distribution is often intractable. Thus, the common strategy is to expand the vector in some other basis and to then do the compression entry-wise (neglecting the dependency between entries of the transformed vector). This is equivalent to do the compression assuming that the signal distribution is the product of the marginal distributions. Thus, if the transformed vector is $\ybold=\Hbold\sbold$, then the normalized redundant information remaining in the compressed signal is
\begin{eqnarray}\label{eq:Rprem}
\text{R}(\Hbold)&=&\frac{1}{N}\left(\text{CSM}(p_{y_1}(y_1)\cdots p_{y_N}(y_N))-\text{MCSM}\right)\nonumber\\&=&\frac{1}{N}\Dbb(p_{\ybold}\left(\ybold\right)\|p_{y_1}(y_1)\cdots p_{y_N}(y_N)),
\end{eqnarray}
where $N$ is the number of entries in $\sbold$. This is the first measure of performance of the transform $\Hbold$ that we are going to use in this paper. Also, this criterion is commonly used in ICA to find the ``most-independent'' representation \cite{Stone2004}.

\subsubsection{Denoising Based on Non-Exact Distribution}\label{sec:ApMSE}
Now, consider the problem of estimating $\sbold$ from the noisy measurement
\begin{eqnarray}\label{eq:ysz}
\zbold=\sbold+\nbold
\end{eqnarray}
where $\nbold$ is an $N$-dimensional white Gaussian noise independent from $\sbold$. Our prior knowledge is the $N$th order pdf $p_{\sbold}(\cdot)$ of the signal. Under these assumptions and according to Stein \cite{Stein1981}, the estimator of minimum mean square error (MMSE) can be represented as
\begin{eqnarray}\label{eq:genmean}
\mathbb{E}\left\{\sbold|\zbold\right\}=\zbold+\sigma^2\nablabold\log{p_\zbold(\zbold)}.
\end{eqnarray}
where $p_\zbold(\zbold)=\left(p_\sbold*p_\nbold\right)(\zbold)$ is the $N$-dimensional pdf of the noisy measurements.
Thus, the MSE given $\zbold$ is 
\begin{eqnarray}\label{eq:genvar}
&~&\mathbb{E}_{\sbold|\zbold}\left\{\left(\sbold-\mathbb{E}\left\{\sbold|\zbold\right\}\right)^2\right\}\nonumber\\&=&\int{\left\|\sbold-\zbold\right\|^2p(\sbold|\zbold)\drm\sbold}-\sigma^4\left\|\nablabold\log{p_\zbold(\zbold)}\right\|^2\nonumber\\&=&N\sigma^2+\sigma^4\Delta\log{p_\zbold(\zbold)}.
\end{eqnarray}
Averaging over $\zbold$, we have
\begin{eqnarray}\label{eq:genMMSE}
\text{MMSE}&=&N\sigma^2-\sigma^4\int{p_{\zbold}(\zbold)\left\|\nablabold\log p_{\zbold}(\zbold)\right\|^2\drm\zbold}\nonumber\\&=&N\sigma^2+\sigma^4\int{p_{\zbold}(\zbold)\Delta\log p_{\zbold}(\zbold)\drm\zbold},
\end{eqnarray}

However, if we perform the MMSE estimator assuming $q_{\sbold}$ as the distribution of $\sbold$, then by using \eqref{eq:genmean}-\eqref{eq:genMMSE}, the MSE of estimation becomes
\begin{eqnarray}\label{eq:genMSE}
\text{MSE}(q_\sbold)=\text{MMSE}+\sigma^4\int{p_\zbold(\zbold)\left\|\nablabold\log\frac{p_\zbold(\zbold)}{q_\zbold(\zbold)}\right\|^2\drm\zbold}
\end{eqnarray}
where $q_\zbold(\zbold)$ is the distribution induced on $\zbold$ in \eqref{eq:ysz} when the distribution on $\sbold$ is $q_\sbold(\sbold)$. Here, notice the pleasing similarity between \eqref{eq:mnb}-\eqref{eq:nb} and \eqref{eq:genMMSE}-\eqref{eq:genMSE}.

If the entries of $\sbold$ are dependent, then the entries of $\zbold$ are dependent, too. Then, performing the exact MMSE estimator is once again often infeasible. The common scheme is then to take $\zbold$ into a transform domain, perform an entry-wise denoising (regardless of the dependency between coefficients), and map the result back into the original domain. If the transformation $\Hbold$ is unitary, the performance of this scheme would be $\text{MSE}(p_{y_1}(y_1)\cdots p_{y_N}(y_N))$ where $\ybold=\Hbold\zbold$ due to Parseval's relation. We write this as a function of  $\Hbold$ normalized by the dimensionality of $\sbold$, with
\begin{eqnarray}\label{eq:criMSE}
\text{MSE}(\Hbold)=\frac{1}{N}\text{MSE}(p_{y_1}(y_1)\cdots p_{y_N}(y_N))
\end{eqnarray}
which is the second measure of performance that we are going to consider in this paper.

\section{Modeling and Wavelet Analysis of S$\alpha$S AR(1) Processes}\label{sec:modeling}
In this section, we first give the definition of continuous-domain S$\alpha$S AR(1) processes and their discrete-domain counterparts. Then, we discuss about the operator-like wavelets that are tuned to this kind of processes.
\subsection{S$\alpha$S AR(1) Processes}\label{subsec:sasproc}
In \cite{Unser-etal2011a}, the authors model the stochastic signal $s$ as a purely innovative process (i.e., \emph{white noise}), having undergone a linear operation. Thus, 
\begin{eqnarray}\label{eq:sw}
s=\Lop^{-1} w
\end{eqnarray}
where $w$ is a continuous-domain white noise and $\Lop^{-1}$ (the inverse of the whitening operator $\Lop$) is a linear operator.

A general white noise is a probability measure on the dual space of a set of test functions that has the following properties \cite{Gelfand-etal1964}: 
\begin{itemize}
\item For a given test function $\varphi$, the statistics of the random variable $\langle w, \varphi\rangle$ do not change upon shifting $\varphi$, where $w$ (the realization of the noise) denotes a generic random element in the dual space of test functions (typically, Schwartz space of tempered distributions).
\item If the test functions in the collection $\{\varphi_\beta\}_{\beta\in B}$ ($B$ is an index set) have disjoint supports, then the random variables in $\{\langle w, \varphi_\beta\rangle\}_{\beta\in B}$ are independent.
\end{itemize}
Under some mild regularity conditions, there is a one-to-one correspondence between the infinitely divisible (id) random variables and the white noises specified above. Thus, specifying a white noise is equivalent to having the random variable $\langle w,\varphi\rangle$ for any test function $\varphi$.

Correspondingly, if $\Lop^*$ denotes the adjoint operator of $\Lop$, then we have
\begin{eqnarray}\label{eq:adjoint}
\langle s,\varphi\rangle=\langle w,\Lop^{-1*}\varphi\rangle
\end{eqnarray}
which means that one can readily deduce the statistical distribution of $\langle s,\varphi\rangle$ from the characterization of the process $w$.

\begin{figure}[t]
\centering
\includegraphics[width=8cm]{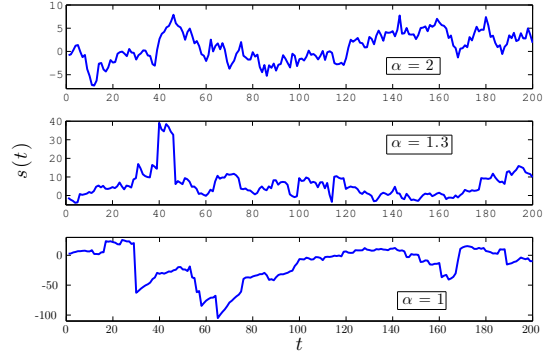}
\caption{Examples of AR(1) processes for different $\alpha$.}\label{fig:AR1s}
\end{figure}

Now, if $w$ is S$\alpha$S white noise, then the random variable $\langle w,\varphi\rangle$ has an S$\alpha$S distribution whose characteristic function is given by
\begin{eqnarray}\label{eq:charfun}
\hat{p}_{\langle w,\varphi\rangle}(\omega)=\mathbb{E}\{\mathrm{e}^{\mathrm{j}\omega\langle w,\varphi\rangle}\}=\mathrm{e}^{-\left|\left\|\varphi\right\|_\alpha\omega\right|^\alpha}.
\end{eqnarray}
In the case of an AR(1) process, we have that
\begin{eqnarray}\label{eq:op}
\Lop = \Dop+\kappa\Iop
\end{eqnarray}
where $\mathrm{D}$ and $\mathrm{I}$ are respectively the differentiator and the identity operator; then, $s$ in \eqref{eq:sw} is a continuous-domain S$\alpha$S AR(1) process. The impulse response of $\Lop^{-1}$ is the causal exponential
\begin{eqnarray}\label{eq:consystem}
\rho_\kappa(t)=\e^{-\kappa t}\Indic_+(t)
\end{eqnarray}
where $\Indic_+(t)$ is the unit step. Thus, as a function of $t$, we can write
\begin{eqnarray}\label{eq:ar1}
s(t)=\left(\rho_\kappa*w\right)(t).
\end{eqnarray}
The AR(1) process is well-defined for $\kappa>0$. The limit case $\kappa=0$ can also be handled by setting the boundary condition $s(0)=0$, which results in a L\'evy process that is non-stationary. Realizations of AR(1) processes for $\kappa=0.05$ and for different values of $\alpha$ are depicted in Figure \ref{fig:AR1s}. When $\alpha$ decreases, the process becomes sparser in the sense that its innovation becomes more and more heavy-tailed.

Now, for a given integer $k$ and time period $T$, set 
\begin{eqnarray}
\varphi_k(t)=\delta(t-kT)-\mathrm{e}^{-\kappa T}\delta\left(t-\left(k-1\right)T\right)
\end{eqnarray}
and define $w_k$ as
\begin{eqnarray}
w_k=\langle s,\varphi_k(t)\rangle=s(kT)-\mathrm{e}^{-\kappa T}s((k-1)T).
\end{eqnarray}
This means that the sampled version $\left\{s_k=s\left((k-1)T\right)\right\}_{k \in \mathbb{Z}}$ of $s(t)$ satisfies the first-order difference equation
\begin{eqnarray}
\label{eq:disAR}
s_{k}=\mathrm{e}^{-\kappa T}s_{k-1}+w_{k}.
\end{eqnarray}
Also, we have that
\begin{eqnarray}
w_k=\langle s,\varphi_k(t)\rangle=\langle w,(\check{\rho_\kappa}*\varphi_k)(t)\rangle
\end{eqnarray}
where $\check{\rho_\kappa}(t)=\rho_\kappa(-t)$ is the impulse response of $\Lop^{-1*}$, the inverse of the adjoint operator of $\Lop$. Also, 
\begin{eqnarray}
(\check{\rho_\kappa}*\varphi_k)(t)=\beta_{\kappa,T}(t-kT)=\Indic_{\big[kT,(k+1)T\big)}\e^{-\kappa (t-kT)}
\end{eqnarray}
is the exponential B-spline with parameters $\kappa$ and $T$ \cite{Unser-etal2011b}. The fundamental property here is that the kernels $\left\{\beta_{\kappa,T}(\cdot-kT)\right\}_{k\in \mathbb{Z}}$ are shifted replicates of each other and have compact and disjoint supports. Thus, according to the definition of a white noise, $\left\{w_k\right\}_{k\in \mathbb{Z}}$ is an iid sequence of S$\alpha$S random variables with the common characteristic function
\begin{eqnarray}\label{eq:charfun}
\hat{p}_w(\omega)=\mathbb{E}\{\mathrm{e}^{\mathrm{j}\omega\langle w,\beta_{\kappa,T}\rangle}\}=\mathrm{e}^{-\left|\|\beta_{\kappa,T}\|_\alpha\omega\right|^\alpha}.
\end{eqnarray}
The conclusion is that a continuous-domain AR(1) process maps into the discrete AR(1) process  $\left\{s_k\right\}_{k\in\mathbb{Z}}$ 
that is uni\-quely specified by \eqref{eq:disAR} and \eqref{eq:charfun}.

We now consider $N$ consecutive samples of the process and define the random vectors $\sbold=\left[s_1~\cdots~s_N\right]^\top$ and $\mathbf{w}=\left[w_1~\cdots~w_N\right]^\top$.
This allows us to rewrite \eqref{eq:disAR} as
\begin{eqnarray}\label{eq:dissLw}
\sbold=\mathbf{L}^{-1}\mathbf{w}
\end{eqnarray}
where $\mathbf{L}^{-1}=\left[\bar{l}_{ij}\right]_{N\times N}$ and
\begin{eqnarray}
\bar{l}_{ij}=\mathrm{e}^{-\kappa T\left(j-i\right)}\cdot \Indic_{\{j\geq i\}}
\end{eqnarray}
which is the discrete-domain counterpart of \eqref{eq:consystem}.

In the next sections, we are going to study linear transforms applied to the signal $s$ (or  $\sbold$). Here, we recall a fundamental property of stable distributions that we shall use in our derivations.

\begin{property}[Linear combination of S$\alpha$S random variables]
\label{prop:sas}
Let $\bar{r}=\sum_{m=1}^M{a_mr_m}$ where $r_m$ are iid S$\alpha$S random variables with dispersion parameter $c$. Then, $\bar{r}$ is an S$\alpha$S as well with dispersion parameter $\left\|[a_1,\cdots,a_M]\right\|_{\alpha}c$.
\end{property}

For more explanation, assume that $r_1, \dots,r_M$ are $M$ iid S$\alpha$S random variables with common characteristic function $\e^{-|c\omega|^\alpha}$, and $a_1,\dots,a_M$ are $M$ arbitrary real numbers. Then, for the characteristic function of the random variable $r^*=\sum_{m=1}^M{a_mr_m}$, we have that
\begin{eqnarray}
\hat{p}_{r^*}(\omega)=\prod_{m=1}^M{\e^{-|a_mc\omega|^\alpha}}=\e^{-\big|\left(\sum_{m=1}^M{\left|a_m\right|^\alpha}\right)^{1/\alpha}c\omega\big|^\alpha}.
\end{eqnarray}
Thus, $r^*$, which is a linear combination of iid S$\alpha$S random variables, is an S$\alpha$S random variable with the same distribution as one of them multiplied by the factor $(\sum_{m=1}^M{|a_m|^\alpha})^{1/\alpha}$; i.e.
\begin{eqnarray}
r^*\,{\buildrel d \over =}\,\Big(\sum_{m=1}^M{|a_m|^\alpha}\Big)^{1/\alpha}r_1.
\end{eqnarray}

\subsection{Operator-Like Wavelets}\label{subsec:OpWT}
\begin{figure}[t]
\begin{minipage}[b]{\linewidth}\label{fig:opwav}
  \centering
\includegraphics[width=8cm]{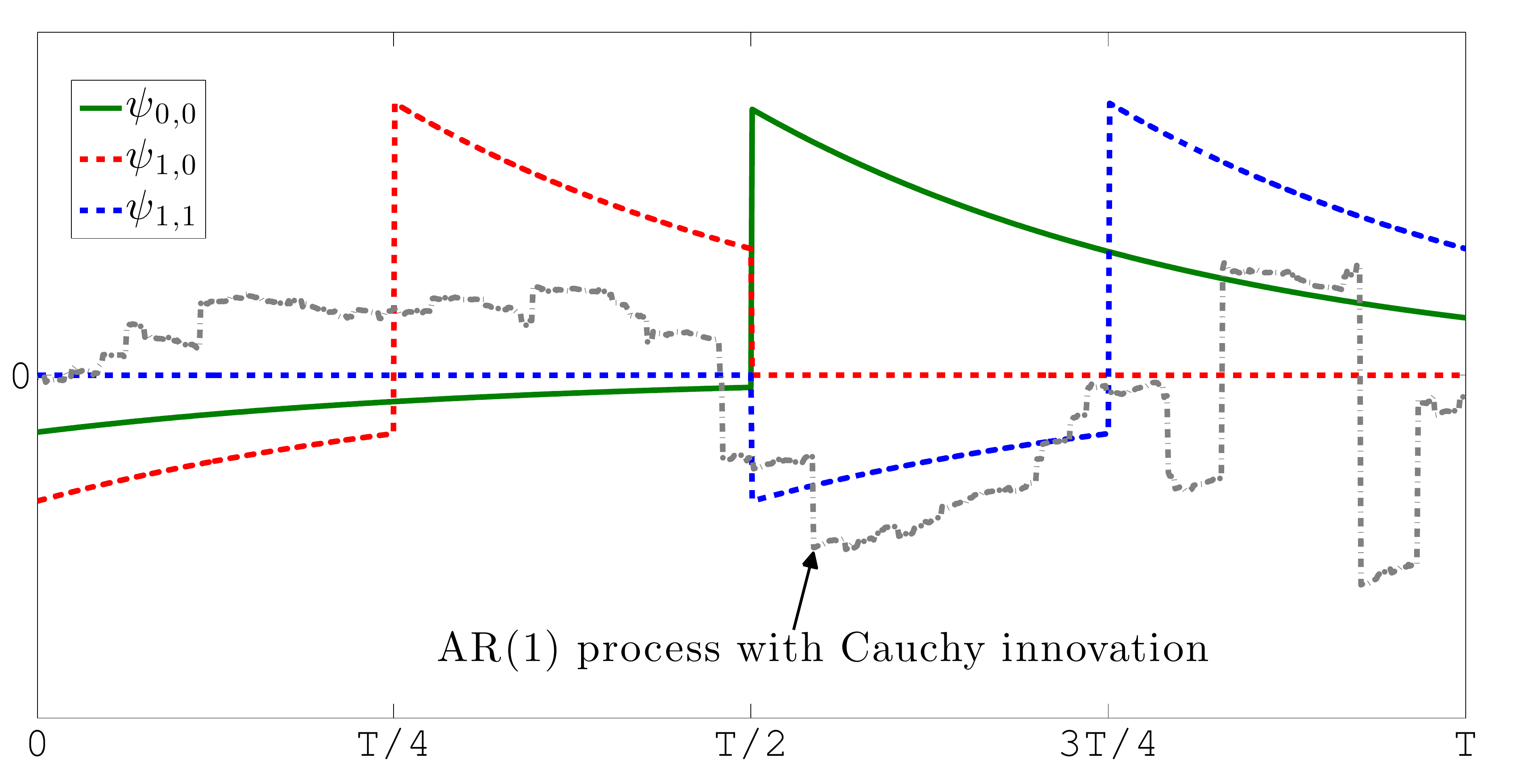}
  \centerline{(a)}
\end{minipage}
\begin{minipage}[b]{\linewidth}
  \centering
\includegraphics[width=8cm]{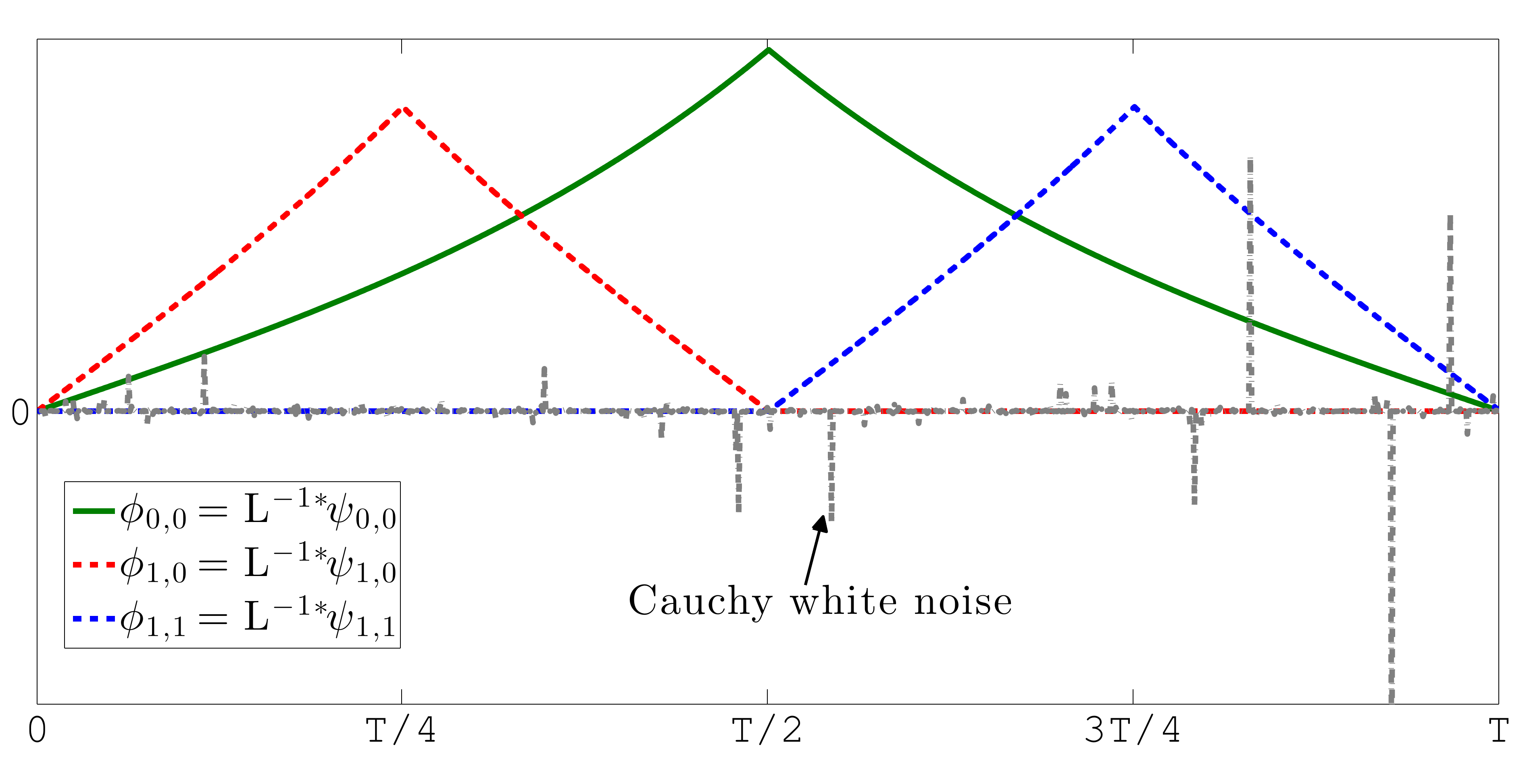}
  \centerline{(b)}
\end{minipage}
\caption{(a) Operator-like wavelets in two consecutive scales acting on an AR(1) process with Cauchy excitation. (b) The equivalent windows (smoothing kernels) acting on the underlying Cauchy white noise. Note that $\psi_{1,0}$ and $\psi_{1,1}$ ($\phi_{1,0}$ and $\phi_{1,1}$, respectively) are non-overlapping.}
\label{fig:lstaropwav}
\end{figure}

Conventional wavelet bases act as smoothed versions of the derivative operator.  To decouple the AR(1) process in \eqref{eq:ar1} by a wavelet-like transform, we need to choose basis functions that essentially behave like the whitening operator $\Lop$ in \eqref{eq:op}. Such wavelet-like basis functions are called operator-like wavelets and can be tailored to any given differential operator $\Lop$ \cite{Khalidov-etal2006}. 
The operator-like wavelet at scale $i$ and location $k$ is given by
\begin{eqnarray}
\psi_{i,k}=\Lop^\ast \phi_i(\cdot-2^i kT),
\end{eqnarray}
where $\phi_i$ is a scale-dependent smoothing kernel. Since $\{\psi_{i,k} \}$ is an orthonormal basis and $s=\Lop^{-1}w$, the wavelet coefficients of the signal $s$ are
\begin{align}\label{eq:lstarwav}
v_{i,k}&=\langle s,\psi_{i,k}\rangle=\langle \Lop^{-1} w,\psi_{i,k}\rangle\\
&=\langle w,\Lop^{-1\ast} \Lop^\ast \phi_i(\cdot-2^i kT)\rangle=\langle w, \phi_i(\cdot-2^i kT)\rangle.\nonumber 
\end{align}
Based on this equality, we understand that, for any given $i$ and for all $k$, the $v_{i,k}$ follows an S$\alpha$S distribution with width parameter  $\|\phi_i\|_\alpha$  \cite{Unser-etal2011a}. Also, since $w$ is independent at every point, intuitively, the level of decoupling has a direct relation to the overlap of the smoothing kernels $\phi_i(\cdot-2^i kT)$.
The operator-like wavelets proposed in \cite{Khalidov-etal2006} are very similar to Haar wavelets, except that they are piecewise exponential
instead of piecewise constant (for $\kappa=0$). Then satisfy
\begin{eqnarray}
&~&\hspace{-.3in}\psi_{i,k}(t)\propto\\\nonumber\\&~&\hspace{-.25in}\e^{-2^{-i}\kappa T}\beta_{\kappa,2^{-i}T}(t-k2^{-i}T)-\beta_{\kappa,2^{-i}T}(t-(k+1)2^{-i}T)\nonumber\\\nonumber\\&~&\hspace{-.3in}=\begin{cases}0 & t<k2^{-i}T \\ \e^{-\kappa (t-(k-1)2^{-i}T)} & k2^{-i}T\leq t<(k+1)2^{-i}T \\ -\e^{-\kappa (t-(k+1)2^{-i}T)} & (k+1)2^{-i}T\leq t<(k+2)2^{-i}T \\ 0 & (k+2)2^{-i}T\leq t\end{cases}\nonumber.
\end{eqnarray}
For these wavelets, the supports of $\phi_{i,k}$ do not overlap within the given scale $i$. Thus, the wavelet coefficients at scale $i$ are independent and identically distributed. This property suggests that this type of transform is an excellent candidate for decoupling AR(1) processes. The illustration of plugging these wavelets into \eqref{eq:lstarwav} is given in Figure \ref{fig:lstaropwav}.

\section{Search for the Optimal Transformation}\label{sec:main}
For now on, we assume that the signal vector $\sbold=\left[s_1~\cdots~s_N\right]^\top$ with $s_k=s\left((k-1)T\right)$ is obtained from the samples of an S$\alpha$S AR(1) process and satisfies the discrete innovation model \eqref{eq:disAR}. The representation of the signal $\sbold$ in \eqref{eq:dissLw} in the transform domain is denoted by $\ybold=[y_1~\cdots~y_N]^\top=\mathbf{H}\sbold$, where $\Hbold=[h_{ij}]_{N\times N}$ is the underlying orthogonal transformation matrix (e.g., DCT, wavelet transform). Let us now use \eqref{eq:Rprem} to characterize the performance of a given transformation matrix $\Hbold$. First, we simplify \eqref{eq:Rprem} to
\begin{eqnarray}
\text{R}(\Hbold)&=&\frac{1}{N}\sum_{n=1}^N{\Hbb(y_n)}-\frac{1}{N}\Hbb(\ybold)\\&=&\frac{1}{N}\sum_{n=1}^N{\Hbb(y_n)}-\Hbb(w_1)-\frac{1}{N}\log\det\Hbold\Lbold^{-1},\nonumber
\end{eqnarray}
where we also observe that $\log\det\Hbold\Lbold^{-1}=0$. In addition, since the $w_m$ is $\alpha$-stable, we can write $y_n\,{\buildrel d \over =}\,\hdot_n w_1$, where $\hdot_n$ is the \mbox{$\alpha$-(pseudo)norm} of the $n$th row of $\Hbold\Lbold^{-1}$
(see Property \ref{prop:sas}) given by
\begin{equation}\label{eq:hbar}
\hdot_n = \left(\sum_{r=1}^N{\left|\sum_{m=1}^N{h_{nm}\bar{l}_{mr}}\right|^\alpha}\right)^{\frac{1}{\alpha}}.
\end{equation}
It follows that
\begin{eqnarray}\label{eq:simRH}
\text{R}(\Hbold)=\frac{1}{N}\sum_{n=1}^N{\log\hdot_n},
\end{eqnarray}
which can be readily calculated for any given $\Hbold$.

\begin{note}
This criterion is reminiscent of the sum-of-dispersion criterion $\sum_{n=1}^N{\hdot_n}$ which is frequently used in the study of $\alpha$-stable stochastic processes \cite{Sahmoudi-etal2005,Soltani-etal1994}. Unlike \eqref{eq:simRH}, the dispersion criterion does not have a direct information-theoretic interpretation. 
\end{note}

As second option, we use the criterion \eqref{eq:criMSE} to measure the performance of a given transform matrix $\Hbold$. Again, it can be simplified to
\begin{eqnarray}\label{eq:mse}
\text{MSE}(\Hbold)=\sigma^2-\frac{\sigma^4}{N}\sum_{n=1}^N{\int{\frac{\left(p'_{y_n}(y_n)\right)^2}{p_{y_n}(y_n)}\mathrm{d}y_n}},
\end{eqnarray}
 in which $y_n$ is the $n$th entry of $\tilde{\ybold}=\Hbold\zbold$. According to Property \ref{prop:sas} of $\alpha$-stable random variables, we write $\tilde{y}_n\,{\buildrel d \over =}\,\hdot_nw_1+n_1$ where $n_1$ is a standard Gaussian random variable. This allows us to deduce the pdf expression
 \begin{eqnarray}\label{eq:pytilde}
 p_{\tilde{y}_n}(y)=\frac{1}{\hdot_n}p_{w_1}\left(\frac{y}{\hdot_n}\right)*p_{n_1}(y).
 \end{eqnarray}
 Thus, \eqref{eq:mse} is calculable through one-dimensional integrals.

For the sake of the optimization process, we also need to derive the gradient of the cost functions $\text{R}$ and $\text{MSE}$ with respect to $\Hbold$. Specifically, according to \eqref{eq:hbar} and \eqref{eq:simRH}, the partial derivative of $\text{R}(\Hbold)$ is
\begin{eqnarray}\label{eq:GradR}
\frac{\partial\text{R}}{\partial h_{ij}}=\frac{1}{N\alpha\hdot_i^\alpha}\frac{\partial\hdot_i^\alpha}{\partial h_{ij}}
\end{eqnarray}
where
\begin{eqnarray}
\frac{\partial\hdot_i^\alpha}{\partial h_{ij}}=\alpha\sum_{r=1}^N{l_{jr}\text{sgn}\left(\sum_{n=1}^N{h_{ik}l_{kr}}\right)\left|\sum_{n=1}^N{h_{ik}l_{kr}}\right|^{\alpha-1}}.
\end{eqnarray}
Also, the partial derivative of $\text{MSE}(\Hbold)$ in \eqref{eq:mse} is
 \begin{eqnarray}\label{eq:PartMSE}
\frac{\partial\text{MSE}}{\partial h_{ij}}&=&-\frac{\sigma^4}{N}\frac{\partial}{\partial\hdot_i}\int{\frac{\left(p^{(1)}_{\tilde{y}_i}(u)\right)^2}{p_{\tilde{y}_i}(u)}\mathrm{d}u}\times\frac{\hdot_i^{1-\alpha}}{\alpha}\frac{\partial\hdot_i^\alpha}{\partial h_{ij}}\nonumber\\\nonumber\\&=&-\frac{\sigma^4}{N}\Bigg(2\int{\frac{\partial}{\partial\hdot_i}p^{(1)}_{\tilde{y}_i}(u)\frac{p^{(1)}_{\tilde{y}_i}(u)}{p_{\tilde{y}_i}(u)}\mathrm{d}u}\nonumber\\&~&~~~~~~-\int{\frac{\partial}{\partial\hdot_i}p_{\tilde{y}_i}(u)\left(\frac{p^{(1)}_{\tilde{y}_i}(u)}{p_{\tilde{y}_i}(u)}\right)^2\mathrm{d}u}\Bigg)\nonumber\\&~&\times~\frac{\hdot_i^{1-\alpha}}{\alpha}\frac{\partial\hdot_i^\alpha}{\partial h_{ij}}
\end{eqnarray}
in which $p^{(k)}_{\tilde{y}_i}(\cdot)$ is the $k$th derivative of $p_{\tilde{y}_i}(\cdot)$ which, according to \eqref{eq:pytilde}, can be written as
\begin{eqnarray}\label{eq:py}
p^{(k)}_{\tilde{y}_i}(\cdot)&=&p_{y_i}(s)*\frac{\drm^k}{\ds^k}\left(\frac{1}{\sqrt{2\pi\sigma^2}}\e^{-\frac{s^2}{2\sigma^2}}\right).
\end{eqnarray}
Also, we have that
\begin{eqnarray}\label{eq:dhpy}
\frac{\partial}{\partial\hdot_i}p_{\tilde{y}_i}(y)=-\frac{1}{\hdot_i}p_{\tilde{y}_i}(y)-\frac{y}{\hdot_i}p^{(1)}_{\tilde{y}_i}(y)-\frac{1}{\hdot_i}p^{(2)}_{\tilde{y}_i}(y)
\end{eqnarray}
and
\begin{eqnarray}\label{eq:dhppy}
\frac{\partial}{\partial \hdot_i}p^{(1)}_{\tilde{y}_i}(y)=-\frac{2}{\hdot_i}p^{(1)}_{\tilde{y}_i}(y)-\frac{y}{\hdot_i}p^{(2)}_{\tilde{y}_i}(y)-\frac{1}{\hdot_i}p^{(3)}_{\tilde{y}_i}(y).
\end{eqnarray}
Now, since the $y_i$ have nice characteristic functions, we can calculate \eqref{eq:py} efficiently through the inverse Fourier transform
\begin{eqnarray}
p^{(k)}_{y_i}(\cdot)=\mathcal{F}_{\omega}^{-1}\left\{(\mathrm{j}\omega)^k\e^{-|\hdot_i\omega|^\alpha-\frac{\sigma^2}{2}\omega^2}\right\}\label{eq:pyder}
\end{eqnarray}
using the FFT algorithm.
 
Thus, we can use gradient-based optimization to obtain the optimal transformations for different values of $\kappa$, $\alpha$, and $N$. For our experiments, we implemented a gradient-descent algorithm with adaptive step size to efficiently find the optimal transform matrix. Since the transform matrix may deviate from the space of unitary matrices, after each step, we project it on that space using the method explained in Appendix \ref{sec:ApProj}. The algorithm is as follows where $C$ is the chosen measure of independence (i.e., $\text{R}$ or $\text{MSE}$):


\begin{algorithm}
\caption{ICA for S$\alpha$S AR(1) Processes}
\label{alg:spatiotemporal}
\begin{algorithmic}[1]\label{alg:opt}
\STATE \textbf{input: } $N, \alpha, \kappa$
\STATE \textbf{initialize: }  $\Hbold_{\text{old}}$, $\mu$, $a\in [1,+\infty)$ and $b\in [0,1]$
\REPEAT
\STATE $\widetilde{\Hbold}_{\text{new}}=\Hbold_{\text{old}}-\mu\nabla C|_{\Hbold_{\text{old}}}$
\STATE Set $\Hbold_{\text{new}}$ to the projection of $\widetilde{\Hbold}_{\text{new}}$ onto the space of unitary matrices
\IF {$C(\Hbold_{\text{new}})<C(\Hbold_{\text{old}})$}
\STATE $\Hbold_{\text{old}}\leftarrow\Hbold_{\text{new}}$
\STATE $\mu\leftarrow a\cdot\mu$
\ELSE
\STATE $\Hbold_{\text{new}}\leftarrow\Hbold_{\text{old}}$
\STATE $\mu\leftarrow b\cdot\mu$
\ENDIF
\UNTIL{ convergence }
\RETURN $\Hbold_{\text{new}}$
\end{algorithmic}
\end{algorithm}

Algorithm \ref{alg:opt} can be viewed as a model-based version of ICA. We take advantage of the underlying stochastic model to derive an optimal solution based on the minimization of \eqref{eq:simRH} and \eqref{eq:mse}, which involves the computation of $\ell_\alpha$-norms of the transformation matrix. By contrast, the classical version of ICA  is usually determined empirically based on the observations of a process, but the ultimate aim is similar; namely, the decoupling of the data vector.

\section{Results for Different Transformations}\label{sec:results}
\begin{figure}[t]
\centering
\includegraphics[width=8.5cm]{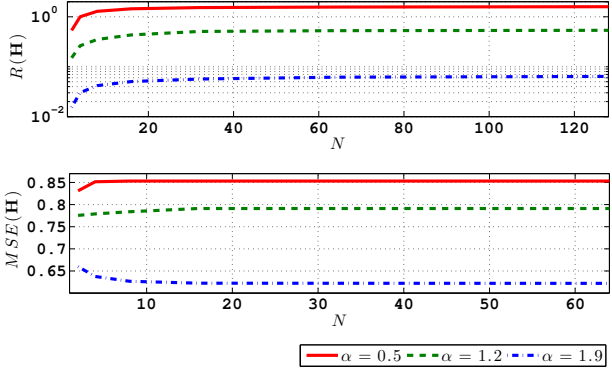}
\caption{Minimum value of $\text{R}(\Hbold)$ and $\text{MSE}(\Hbold)$ for L\'evy processes as a function of $N$ for different values of $\alpha$. In the second plot $\sigma^2=1$.}\label{fig:CodLevyN}
\end{figure}

Initially, we investigate the effect of the signal length $N$ on the value of $\text{R}$ and $\text{MSE}$. We consider the case of a L\'evy process (i.e., $\kappa=0$) and numerically optimized the criteria for different $\alpha$ and plot it as a function of $N$. Results are depicted in Figure \ref{fig:CodLevyN}. As we see, the criteria values converge quickly to their asymptotic values. Thus, for the remainder of the experiments, we have chosen $N=64$. This is a block size that is reasonable computationally and large enough to be representative of the asymptotic regime.

\begin{figure}[t]
\centering
\includegraphics[width=8cm]{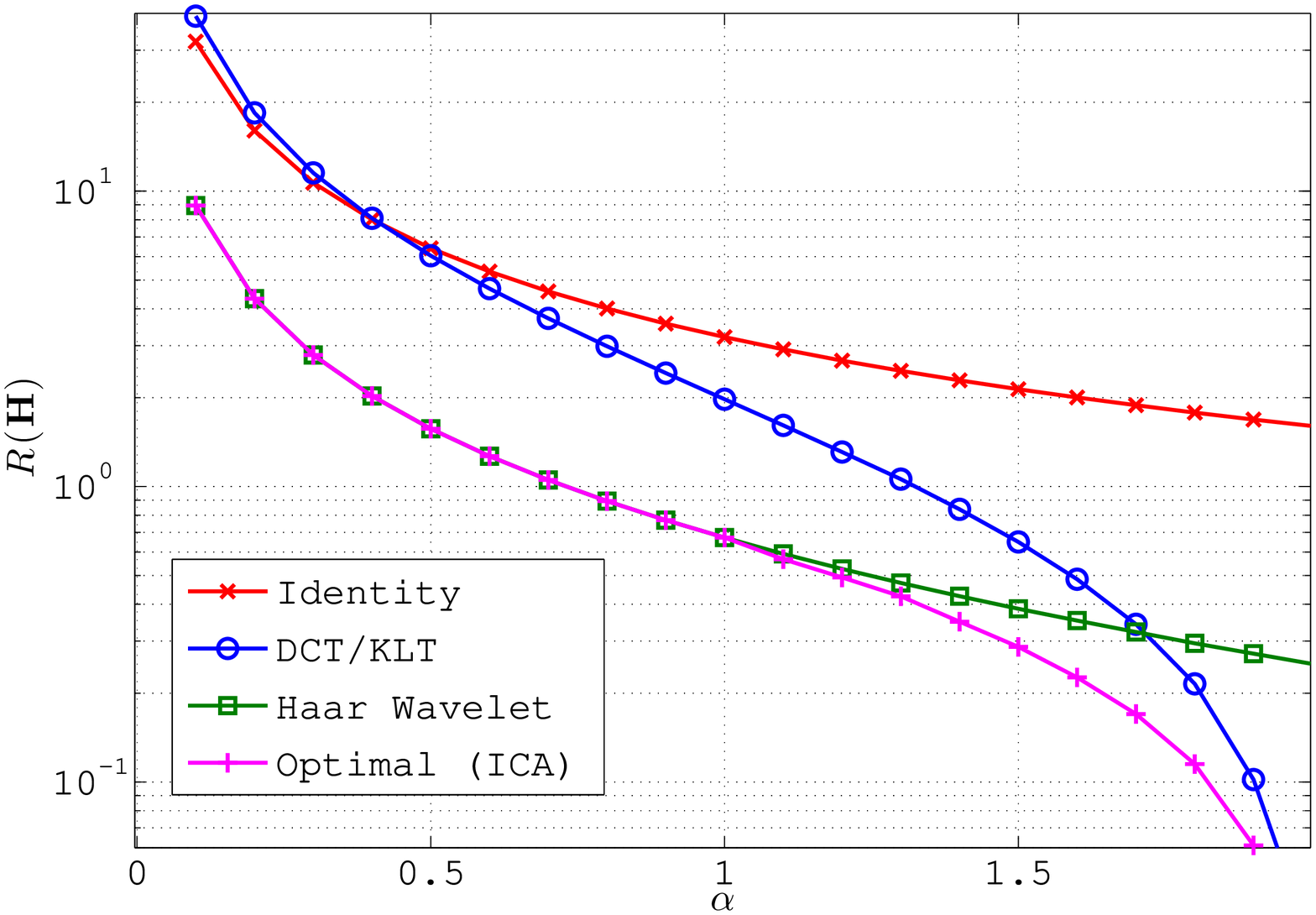}
\caption{$\text{R}(\Hbold)$ of L\'evy processes versus $\alpha$ when $n=64$ for different $\Hbold$.}\label{fig:CodLevy}
\end{figure}

\begin{figure}[t]
\centering
\includegraphics[width=8cm]{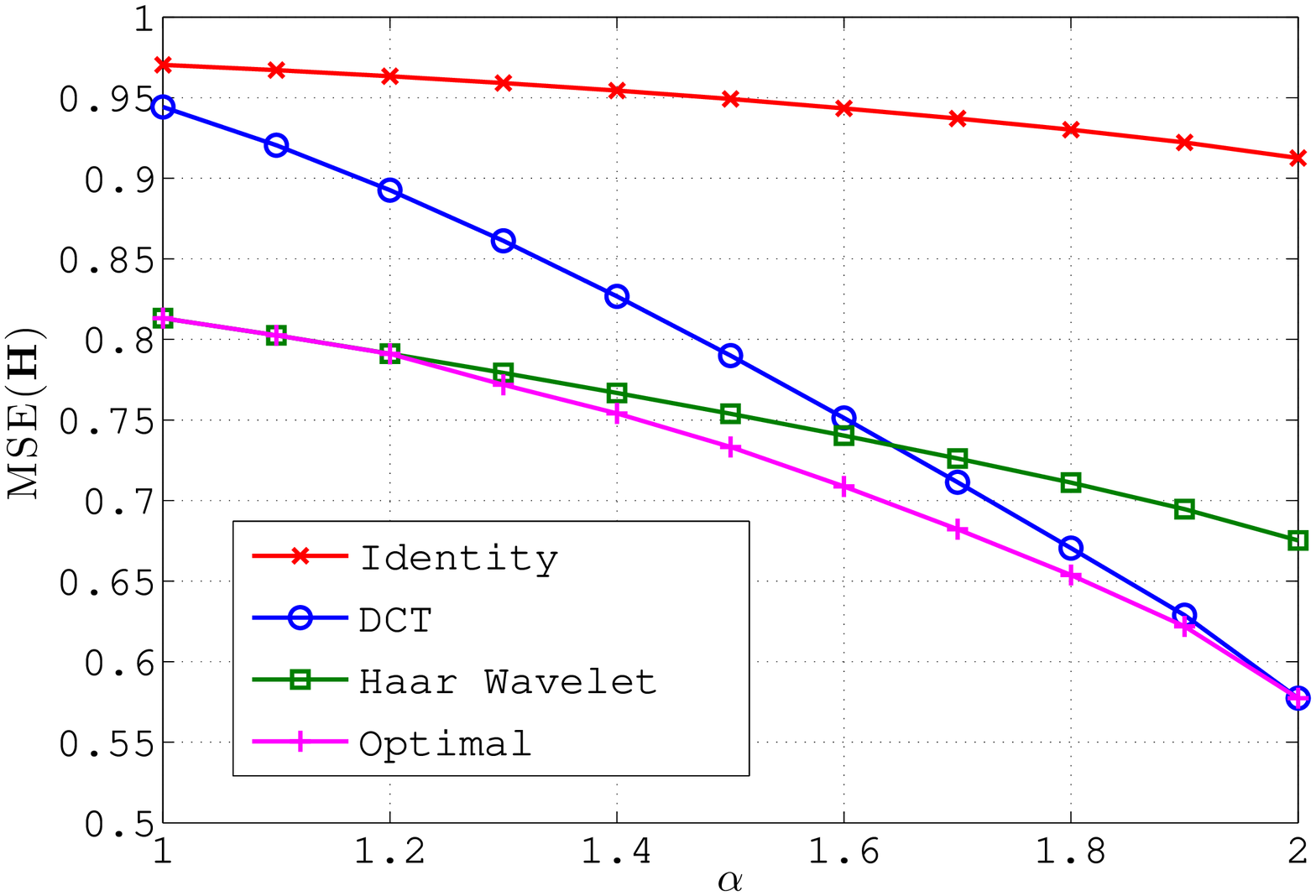}
\caption{$\text{MSE}(\Hbold)$ of L\'evy processes versus $\alpha$ when $n=64$ for different $\Hbold$.}\label{fig:DenLevy}
\end{figure}

Then, we investigate the performance of different transforms for various processes. First, we focus on the L\'evy processes. In this case, the operator-like wavelet transform is the classical Haar wavelet transform (HWT). The performance criteria $\text{R}$ and $\text{MSE}$ as a function of $\alpha$ for various transforms are plotted in Figures \ref{fig:CodLevy} and \ref{fig:DenLevy}, respectively. The considered transformations are as follows: identity as the baseline, discrete cosine transform (DCT), Haar wavelet transform (HWT), and optimal solution (ICA) provided by the proposed algorithm. In the case of $\alpha=2$ (Gaussian scenario), the process $s$ is a Brownian motion whose KLT is a sinusoidal transform that is known analytically.  In this case, the DCT and the optimal transform converge to the KLT since being decorrelated is equivalent to being independent. We see this coincidence in both Figures \ref{fig:CodLevy} and \ref{fig:DenLevy}. The vanishing of $\text{R}$ at $\alpha=2$ indicates perfect decoupling. By contrast, as $\alpha$ decreases, neither the DCT nor the optimal transform decouples the signal completely. The latter means that there is no unitary transform that completely decouples stable non-Gaussian L\'evy processes. However, we see that, based on both criteria $\text{R}$ and $\text{MSE}$, and as $\alpha$ decreases, the DCT becomes less favorable while the performance of the HWT gets closer to the optimal one. Moreover, Figures \ref{fig:CodLevy} and \ref{fig:DenLevy} even suggest that the Haar wavelet transform is equivalent to the ICA solution for $\alpha\leq1$.

\begin{figure*}[t]
\centering
\includegraphics[width=17cm]{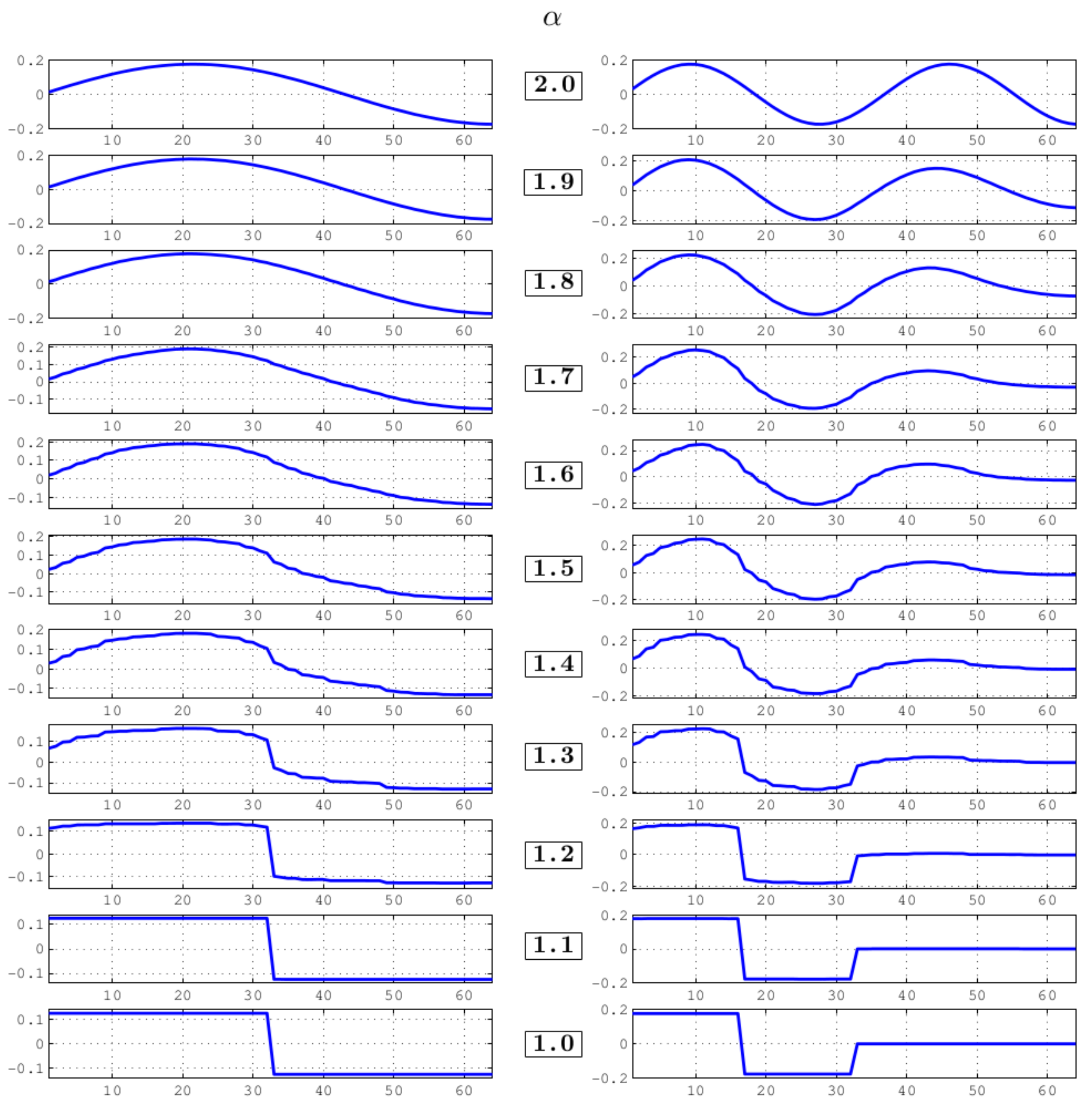}
\caption{Two rows of the optimal $\Hbold$ (ICA) for $\alpha=2$ down to $1$ when $N=64$. In each row, we see the evolution from sinusoidal waves to Haar wavelets by increasing the sparsity of the underlying innovation process.}\label{fig:Evol}
\end{figure*}

Also, to see the transition from sinusoidal bases to Haar wavelet bases, we plot the optimal basis which is obtained by the proposed algorithm at two consequent scales. In Figure \ref{fig:Evol}, we see the progressive evolution of the ICA solution from the sinusoidal basis to the Haar basis while changing the parameter $\alpha$ of the model.

\begin{figure}[t]
\centering
\includegraphics[width=8cm]{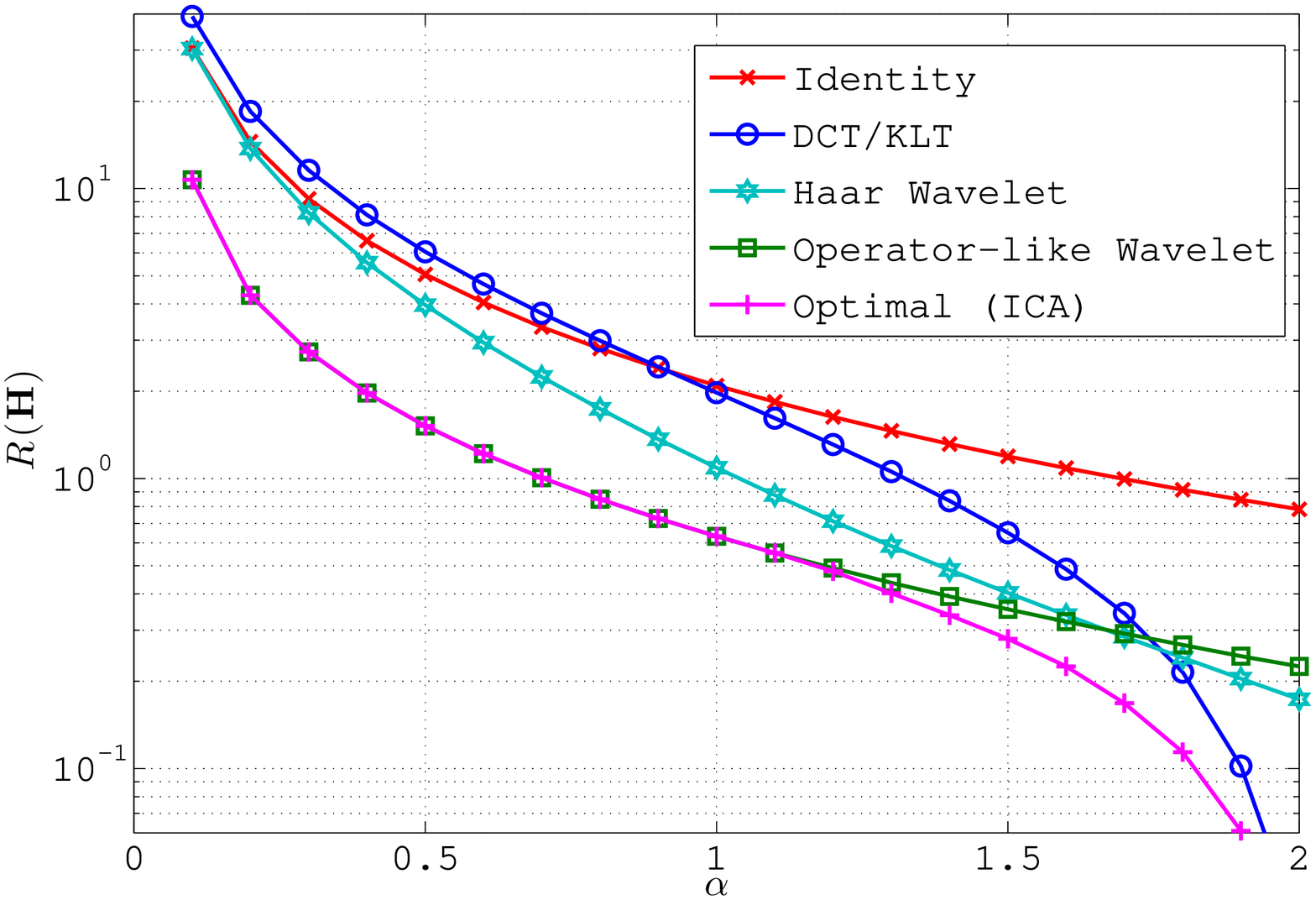}
\caption{$\text{R}(\Hbold)$ versus $\alpha$ when $\mathrm{e}^{-\kappa T}=0.9$ and $n=64$ for different $\Hbold$.}\label{fig:CodAr1}
\end{figure}

\begin{figure}[t]
\centering
\includegraphics[width=8cm]{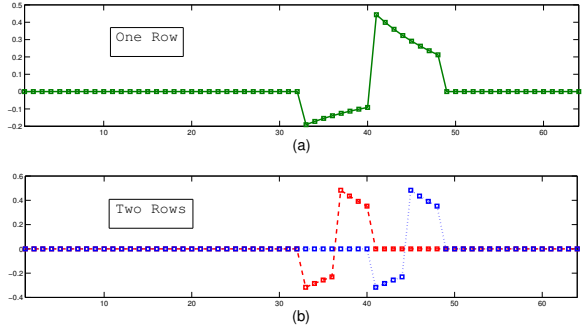}
\caption{Three rows of the optimal $\Hbold$ for $\alpha=1$ and $n=64$. Parts (a) and (b) show the dyadic structure of the wavelets.}\label{fig:Rows}
\end{figure}

Next, we consider a stationary AR(1) process with $\mathrm{e}^{-\kappa T}=0.9$ and $n=64$. For $\alpha=2$, we get the well-known classical Gaussian AR(1) process for which the DCT is known to be asymptotically optimal \cite{Pearl1973,Unser1984}. The performance criterion $\text{R}$ versus $\alpha$ for the DCT, the HWT, the operator-like wavelet matched to the process, and the optimal ICA solution are plotted in Figure \ref{fig:CodAr1}. Here too we see that, for $\alpha=2$, ICA is equivalent the DCT. But, as $\alpha$ decreases, the DCT loses its optimality and the matched operator-like wavelet becomes closer to optimum. Again, we observe that, for $\alpha\leq1$, the ICA solution is the matched operator-like wavelet described in Section \ref{subsec:OpWT}. The fact that the matched operator-like wavelet outperforms the HWT shows the benefit of the tuning of the wavelet to the differential characteristics of the process. Also, as shown in Figure \ref{fig:Rows}, experimentally determined ICA basis functions for $\alpha=1$ are indistinguishable from the wavelets in Figure \ref{fig:lstaropwav}.

To substantiate those findings, we present a theorem that states that, based on the above mentioned criteria and for any $\alpha<2$, the operator-like wavelet transform outperforms the DCT (or, equivalently, the KLT associated with the Gaussian member of the family) as the block-size $N$ tends to infinity.
\begin{theorem}\label{th:r}
If $\alpha<2$ and $\kappa\geq 0$, we have that
\begin{eqnarray}\label{eq:thpart1}
\lim_{N\rightarrow\infty}\text{R}(\text{OpWT})<\lim_{N\rightarrow\infty}\text{R}(\text{DCT})=\infty
\end{eqnarray}
and
\begin{eqnarray}\label{eq:thpart2}
\lim_{N\rightarrow\infty}\text{MSE}(\text{OpWT})<\lim_{N\rightarrow\infty}\text{MSE}(\text{DCT})=\sigma^2,
\end{eqnarray}
where OpWT stands for the operator-like wavelet transform.
\end{theorem}
The proof is given in Appendix \ref{sec:approof}.

In addition, this theorem states that, for $\alpha<2$ and as $N$ tends to $\infty$, the performance of the DCT is equivalent to the trivial identity operator. This is surprising because, since the DCT is optimal for the Gaussian case ($\alpha=2$), one may expect that it has a good result for other AR(1) processes. However, although this theorem does not assert that operator-like wavelets are the optimal basis, it still shows that, by applying them, we obtain a better performance than trivial transformations. Also, through simulations we observed that operator-like wavelets are close to optimal transform, particularly when the underlying white noise becomes very sparse.

\section{Summary and Future Studies}
In this paper, we focused on the simplest version (first-order differential system with an S$\alpha$S excitation) of the sparse stochastic processes which have been proposed by Unser et al \cite{Unser-etal2011a,Unser-etal2011b}. Because of the underlying innovation model and the properties of S$\alpha$S random variables, we could obtain a closed-form formula for the performance of different transform-domain representations and characterize the optimal transform. This is a novel model-based point of view for ICA. We proved that operator-like wavelets are better than sinusoidal transforms for decoupling the AR(1) processes with sparse excitations ($\alpha<2$). This result is remarkable since sinusoidal bases are known to be asymptotically optimal for the classical case of $\alpha=2$. Moreover, we showed that, for very sparse excitations ($\alpha\lesssim 1$), operator-like wavelets are equivalent to the ICA. As far as we know, this is the first theoretical results on the optimality of wavelet-like bases for a given class of stochastic processes.

Another interesting aspect of this study is that it gives a unified framework for Fourier-type transforms and a class of wavelet transforms. Now, the Fourier transform and the wavelet transforms were based on two different intuitions and philosophies. However, here we have a model in which we obtain both transform families just by changing the underlying parameters.

The next step in this line of research is to investigate the extent to which these findings can be generalized to other white noises or higher-order differential operators. Also, studying the problem in the original continuous domain would be theoretically very valuable. 
 
\appendices
\section{Projection on the Space of Unitary Matrices}\label{sec:ApProj}
Suppose that $\Abold$ is an $N\times N$ matrix. Our goal is to find the unitary matrix $\Hbold^*$ that is the closest to $\Abold$ in Frobenius norm, in the sense that
\begin{eqnarray}
\Hbold^*=\text{arg}\min_{\Hbold}\|\Abold-\Hbold\|_F.
\end{eqnarray}
According to singular-value decomposition (SVD), we can write $\Abold=\Ubold\Lambdabold\Vbold^\top$ where $\Ubold$ and $\Vbold$ are unitary matrices and $\Lambdabold$ is a diagonal matrix with nonnegative diagonal entries.

Since the Frobenius norm is unitarily invariant, we have that
\begin{eqnarray}\label{eq:fer}
\|\Abold-\Hbold\|_F=\|\Lambdabold-\Ubold^\top\Hbold\Vbold\|_F
\end{eqnarray}
in which $\Ubold^\top\Hbold\Vbold$ is a unitary matrix that we call $\Kbold$. The expansion of the right-hand side of \eqref{eq:fer} gives
\begin{eqnarray}\label{eq:lambdak}
\|\Lambdabold-\Kbold\|_F^2&=&\sum_{1\leq i,j\leq N}{k_{ij}^2}+\sum_{i=1}^N{\lambda_{ii}^2}-2\sum_{i=1}^N{\lambda_{ii}k_{ii}}\nonumber\\&=&N+\sum_{i=1}^N{\lambda_{ii}^2}-2\sum_{i=1}^N{\lambda_{ii}k_{ii}}.
\end{eqnarray}
Since $\Kbold$ is unitary, $|k_{ii}|\leq 1$ for $i=1,\dots,N$. Thus, setting $k_{ii}=1$, which means setting $\Kbold=\Ibold$, minimizes \eqref{eq:lambdak}. Consequently, the projection of $\Abold$ on the space of unitary matrices is $\Hbold^*=\Ubold\Vbold^\top$.

\section{Proof of Theorem \ref{th:r}}\label{sec:approof}
\subsection{Proof of Part 1 (Equation \eqref{eq:thpart1})}
According to \eqref{eq:simRH}, we have that
\begin{eqnarray}\label{eq:Rempirical}
\text{R}(\Hbold)&=&\frac{1}{N}\sum_{n=1}^N{\log\hdot_n}=\frac{1}{N}\sum_{n=1}^N{\log\left(\frac{1}{\hdot_n^{-1}}\right)}\nonumber\\&=&\int_{\mathbb{R}}{\log\left(\frac{1}{\gamma}\right)p\left(\gamma\right)\mathrm{d}\gamma}
\end{eqnarray}
in which $p(\cdot)$ is the empirical distribution of $\hdot_n^{-1}$.

According to SVD, we can write $\Lbold^{-1}=\Ubold\Lambdabold\Vbold^\top$ where $\Lambdabold$ is a diagonal matrix with $\lambda_i$ as diagonal entries. Taking $\sbold$ in the KLT domain is equivalent to multiplying it by $\Ubold^\top$. The eigenvalues of the covariance of AR(1) matrices are known in closed form and are given by \cite{Ray-etal1970} and \cite{Kamilov-etal2012}, for $\kappa\geq 0$, as
\begin{eqnarray}\label{eq:lambdas}
|\lambda_i|^{-1}=\sqrt{\left(1-\e^{-\kappa T}\right)^2+4\e^{-\kappa T}\sin^2\left(\frac{\omega_i}{2}\right)}
\end{eqnarray}
and
\begin{eqnarray}
v_{ij}=\sqrt{\frac{2}{N+\left(1-\e^{-2\kappa T}\right)\lambda^{2}_i}}\nonumber\\&~&\hspace{-1.1in}\times\sin\left(\omega_i\left(j-\frac{N+1}{2}\right)+i\frac{\pi}{2}\right)
\end{eqnarray}
in which $\omega_i$, $i=1,\dots,N$, is the $i$th positive root of 
\begin{eqnarray}\label{eq:omegas}
\tan{(N\omega)}=-\frac{\left(1-\e^{-2\kappa T}\right)\sin{\omega}}{\cos{\omega}-2\e^{-\kappa T}+\e^{-2\kappa T}\cos{\omega}}.
\end{eqnarray}
Since $\tan{(N\omega)}$ is an injective function that sweeps the whole domain of the real numbers while $\omega\in\left[\frac{i-1}{N}\pi,\frac{i}{N}\pi\right]$, for $i=1,\dots,N$, \eqref{eq:omegas} has a single root in each of such intervals. Thus, as $N$ tends to infinity, the empirical distribution of the $\omega_i$ tends to the uniform distribution on $\left[0,\pi\right]$. Then, starting from \eqref{eq:lambdas}, one can obtain the limit empirical distribution of $|\lambda_i|$ as
\begin{eqnarray}
p_{\lambda}(\lambda)=\frac{2}{\pi}\frac{\lambda}{\sqrt{\lambda^2-\left(1-\e^{-\kappa T}\right)^2}\sqrt{\left(1+\e^{-\kappa T}\right)^2-\lambda^2}}.
\end{eqnarray}
Now, $\sum_{j=1}^Nv_{ij}^2=1$ means that
\begin{eqnarray}
\sum_{j=1}^N\left|\sin\left(\omega_i\left(j-\frac{N+1}{2}\right)+i\frac{\pi}{2}\right)\right|^2 \sim\mathcal{O}\left(N\right)
\end{eqnarray}
 as $N$ tends to infinity. But, for $\alpha<2$, we have that
 \begin{eqnarray}
&~&\hspace{-0.15in}\Bigg(\sum_{j=1}^N\left|\sin\left(\omega_i\left(j-\frac{N+1}{2}\right)+i\frac{\pi}{2}\right)\right|^\alpha\Bigg)^{\frac{1}{\alpha}}\\&~&\hspace{-0.3in}\geq\Bigg(\sum_{j=1}^N\left|\sin\left(\omega_i\left(j-\frac{N+1}{2}\right)+i\frac{\pi}{2}\right)\right|^2\Bigg)^{\frac{1}{\alpha}}\sim\mathcal{O}(N^\frac{1}{\alpha}).\nonumber
\end{eqnarray}
Thus, for $\alpha<2$, $\big(\sum_{j=1}^N\left|v_{ij}\right|^\alpha\big)^{\frac{1}{\alpha}}$ grows faster than $\mathcal{O}(N^{\frac{1}{\alpha}-\frac{1}{2}})$ and thus tends to infinity as $N$ tends to infinity.
Consequently, the limit empirical distribution of $\hdot_i^{-1}$ can be represented as
\begin{equation}\label{eq:gamdist}
p(\gamma)=\begin{cases}\frac{2}{\pi}\frac{\gamma}{\sqrt{\gamma^2-\left(1-\e^{-\kappa T}\right)^2}\sqrt{\left(1+\e^{-\kappa T}\right)^2-\gamma^2}} & \alpha=2 \\ \delta(\gamma) & \alpha\neq 2.\end{cases}
\end{equation}
By plugging this result into \eqref{eq:Rempirical}, we conclude that, for $\alpha<2$, $\lim_{N\rightarrow\infty}\text{R}(\text{KLT})=\infty$. This completes the proof of the right-hand side.

Now, for the proof of the left-hand side, we need to specify the matrix $\Hbold$ for the operator-like wavelet transform. This matrix is given by the recursive construction
\begin{eqnarray}
&~&\hspace{-.3in}\Hbold_k=\text{diag}\Bigg(\sqrt{\frac{1-\e^{-2\kappa T}}{1-\e^{-2^{k+1}\kappa T}}},\sqrt{\frac{1-\e^{-2\kappa T}}{1-\e^{-2^{k+1}\kappa T}}},\overbrace{1,\dots,1}^{2^k-2}\Bigg)\nonumber\\&~&\hspace{0.2in}\times\left[\begin{matrix}\mathbf{\ell}_{k-1} & \e^{-2^{k-1}\kappa T}\mathbf{\ell}_{k-1}\\ - \e^{-2^{k-1}\kappa T}\mathbf{\ell}_{k-1}&\mathbf{\ell}_{k-1} \\ \Hbold'_{k-1} & \mathbf{0} \\ \mathbf{0} & \Hbold'_{k-1}\end{matrix}\right]
\end{eqnarray}
in which $\Hbold'_{k-1}$ is the matrix $\Hbold_{k-1}$ omitting the first row and $\mathbf{\ell}_{k-1}=[1,\e^{-\kappa T},\dots,\e^{-(2^{k-1}-1)\kappa T}]$. Also, $\Hbold_0=[1]$. Let us denote the empirical distribution of $\hdot_i^{-1}$ (the reciprocal of the $\alpha$-(pseudo) norm of the rows of $\Hbold_k\Lbold_{2^k}$) by $p_k(\gamma)=\sum_{i=1}^{k}{p_i\delta(\gamma-\gamma_i)}$. Now, for the sequence of $p_i$ and $\gamma_i$, with respect to $k$, we have the following recursive relation:
\begin{itemize}
\item Replace $p_{k-1}$ by $\left(\frac{p_{k-1}}{2},\frac{p_{k-1}}{2}\right)$
\item Remove $\gamma_{k-1}$. Then, if $\kappa>0$, set
\begin{eqnarray}\label{eq:gkar}
&~&\hspace{-0.3in}\gamma_{k-1}=\sqrt{\frac{1-\e^{-2^{k+1}\kappa T}}{1-\e^{-2\kappa T}}}\\&~&\times\Bigg(\sum_{i=-2^{k-1}+1}^{2^{k-1}}{\left(\frac{\e^{-|i|\kappa T}-\e^{-(2^k-|i|)\kappa T}}{1-\e^{-2\kappa T}}\right)^\alpha}\Bigg)^{-\frac{1}{\alpha}}\nonumber
\end{eqnarray}
and
\begin{eqnarray}
\gamma_k=\sqrt{\frac{1-\e^{-2^{k+1}\kappa T}}{1-\e^{-2\kappa T}}}\Bigg(\sum_{i=1}^{2^k}{\left(\frac{1-\e^{-2i\kappa T}}{1-\e^{-2\kappa T}}\right)^\alpha}\Bigg)^{-\frac{1}{\alpha}}
\end{eqnarray}
else, if $\kappa=0$, set
\begin{eqnarray}
\gamma_{k-1}=2^{\frac{k}{2}}\Bigg(\sum_{i=-2^{k-1}+1}^{2^{k-1}}{\left(2^{k-1}-\left|i\right|\right)^\alpha}\Bigg)^{-\frac{1}{\alpha}}
\end{eqnarray}
and
\begin{eqnarray}\label{eq:gklevy}
\gamma_k=2^{\frac{k}{2}}\Bigg(\sum_{i=1}^{2^{k}}{i^\alpha}\Bigg)^{-\frac{1}{\alpha}}.
\end{eqnarray}
\end{itemize}

Consequently, according to \eqref{eq:Rempirical}, we have that
\begin{eqnarray}\label{eq:redundsum}
\lim_{n\rightarrow\infty}\text{R}(\text{HWT})=\sum_{k=1}^\infty{2^{-k}\log\gamma_k^{-1}}.
\end{eqnarray}
However, for the case $\kappa>0$ and $k<N$, 
\begin{eqnarray}
\gamma_k^{-1}&\leq&\frac{2\left(\left(2^k-1\right)\left(1-\e^{-2^k\kappa T}\right)^{\alpha}\right)^{\frac{1}{\alpha}}}{\sqrt{\left(1-\e^{-2\kappa T}\right)\left(1-\e^{-2^{k+1}\kappa T}\right)}}\nonumber\\&\leq&\frac{2}{\sqrt{1-\e^{-2\kappa T}}}\sqrt{\frac{1-\e^{-2^{k}\kappa T}}{1+\e^{-2^{k}\kappa T}}}\left(2^k-1\right)^{\frac{1}{\alpha}}\nonumber\\&\leq&\frac{2^{1+\frac{k}{\alpha}}}{\sqrt{1-\e^{-2\kappa T}}}.
\end{eqnarray}
Thus,
\begin{eqnarray}
\lim_{n\rightarrow\infty}\text{R}(\text{HWT})&\leq&\sum_{k=1}^\infty{2^{-k}\log{\frac{2^{1+\frac{k}{\alpha}}}{\sqrt{1-\e^{-2\kappa T}}}}}\nonumber\\&=&\Big(\frac{2}{\alpha}+\frac{1}{2}\log{\frac{1}{1-\e^{-2\kappa T}}}\Big)\log 2.
\end{eqnarray}
For the case $\kappa=0$ and $k<N$,
\begin{eqnarray}
\gamma_k^{-1}&\leq&2^{-\frac{k}{2}}\left(\left(2^k-1\right)\left(2^{k-1}\right)^{\alpha}\right)^{\frac{1}{\alpha}}\nonumber\\&\leq&2^{\frac{k}{2}+\frac{k}{\alpha}-1}.
\end{eqnarray}
Thus,
\begin{eqnarray}
\lim_{n\rightarrow\infty}\text{R}(\text{HWT})\leq\sum_{k=1}^\infty{2^{-k}\log{2^{\frac{k}{2}+\frac{k}{\alpha}-1}}}=\frac{2}{\alpha}\log 2.
\end{eqnarray}
Therefore, the proof is complete.

\subsection{Proof of Part 2 (Equation \eqref{eq:thpart2})}\label{sec:approof2}
\emph{Proof: } We have that
\begin{equation}\label{eq:msedistr}
\text{MSE}(\Hbold)=\frac{1}{N}\sum_{n=1}^N{\nu(\hdot_n^{-1})}=\int_{\mathbb{R}}{\nu(\gamma^{-1})p\left(\gamma\right)\mathrm{d}\gamma}
\end{equation}
in which $\nu(\gamma^{-1})$ is the MMSE of the estimating $w$ from $s$ in the scalar problem 
\begin{eqnarray}\label{eq:scalMSE}
s=\gamma^{-1} w+z,
\end{eqnarray}
where $w$ is a stable random variable with characteristic function $\hat{p}_w(\omega)=\exp\left(-|\omega|^\alpha\right)$ and $z$ is a Gaussian random variable with variance $\sigma^2$. We know that $\nu(\cdot)$ is a monotone continuous function that vanishes at zero and tends to $\sigma^2$ asymptotically. Also, $p(\cdot)$ is the empirical distribution of the reciprocals of $\hdot_i$ in \eqref{eq:hbar}. The proof is then essentially the same as the one of Theorem \ref{th:r} but simpler since the function $\nu(\cdot)$ is bounded.

 For $\Hbold$ equal to Fourier transform, the limiting $p(\gamma)$ was given in \eqref{eq:gamdist}. Thus, for $\alpha<2$, as $n$ tends to infinity, $\text{MSE}(\Hbold)$ tends to $\sigma^2$. This completes the proof of the right-hand side.
 
 For the case that $\Hbold$ is the operator-like wavelet transform, the limmit is $p(\gamma)=\sum_{k=1}^\infty{p_k\delta\left(\gamma-\gamma_k\right)}$ where $p_k=2^{-k}$ and $\gamma_k$ were given in \eqref{eq:gkar} -- \eqref{eq:gklevy}. Thus, we have that
\begin{eqnarray}
\text{MSE}(\text{OpWT})=\sum_{k=1}^\infty{2^{-k}\nu(\gamma_k^{-1})}\leq\frac{1}{2}\nu(\gamma_1^{-1})+\frac{\sigma^2}{2}.
\end{eqnarray} 
But, obviously, $\gamma_1^{-1}<\infty$; hence, $\nu(\gamma_1^{-1})<\sigma^2$, which completes the proof.

\bibliographystyle{IEEEtran}
\bibliography{MinimumDependence.bbl}

\end{document}